\documentclass{emulateapj}

\usepackage{float}
\usepackage{amsmath}
\usepackage{epsfig,floatflt}
\usepackage{subfigure}

\newcommand{\mincir}{\raise
-2.truept\hbox{\rlap{\hbox{$\sim$}}\raise5.truept
\hbox{$<$}\ }}
\newcommand{\magcir}{\raise
-2.truept\hbox{\rlap{\hbox{$\sim$}}\raise5.truept
\hbox{$>$}\ }}
\newcommand{\minmag}{\raise-2.truept\hbox{\rlap{\hbox{$<$}}\raise
6.truept\hbox{$>$}\ }}
\newlength{\royalength}
\newcommand{\be}{\begin{equation}}
\newcommand{\ee}{\end{equation}}
\newcommand{\ba}{\begin{eqnarray}}
\newcommand{\ea}{\end{eqnarray}}
\newcommand{\brr}{\begin{array}}

\newcommand{\err}{\end{array}}
\newcommand{\bc}{\begin{center}}
\newcommand{\ec}{\end{center}}

\begin{document}

\title{Dark matter caustics and the enhancement of self-annihilation flux}

\author{Roya C. Mohayaee\altaffilmark{1}, Sergei Shandarin\altaffilmark{2}, 
Joseph Silk\altaffilmark{3}}

\altaffiltext{1}{Institut d'astrophysique de Paris, 98 bis boulevard Arago, France}

\altaffiltext{2}{Department of Physics and Astronomy, University of Kansas, KS 66045, U.S.A.}

\altaffiltext{3}{University of Oxford, Astrophysics, Keble Road, Oxford OX1 3RH, U.K.}

\begin{abstract}

Cold dark matter haloes are populated by caustics, which are
yet to be resolved in N-body simulations or observed in the Universe. 
Secondary infall model provides a paradigm for the study of
caustics in {\it typical} haloes assuming that they have had no major 
mergers and have grown only by smooth accretion. This is a particular 
characteristic of the smallest dark matter 
haloes of about $10^{-5}$ M$_\odot$, which although {\it atypical} contain no
substructures and could have survived until now with no major mergers.
Thus using this model as the first guidline,
we evaluate the neutralino self-annihilation flux for these haloes. 
Our results show that caustics could leave a distinct sawteeth signature on the
differential and cumulative fluxes coming from the outer regions of these haloes. 
The total annihilation signal from the regions away 
from the centre can be boosted by about forty percents. 
\end{abstract}


\maketitle


\section{Introduction}
Evidence from the
rotation curves of  galaxies, gravitational lensing, microwave background
radiation, peculiar velocity fields, and many other observations
indicate that the visible mass, in the form of stars and hot gas, is only a 
small fraction of the total content of the Universe.
The nature of the missing mass, the {\it dark matter}, remains unknown but is
widely presumed to be Weakly Interacting Massive Particles (WIMPs), 
such as the lightest  supersymmetric particles,  which 
are yet to be detected in particle accelerators (Jungman, Kamionkowski,
\& Griest 1996 ; Bertone, Hooper \& Silk 2004). 

Accelerator searches are complemented by the vast experimental efforts to 
detect these particles in our galaxy and in  nearby galaxies which are
believed to be embedded in dark matter haloes (Ostriker \& Peebles 1973).
Such complementary techniques presently involve direct detection in 
low background laboratory detectors (Goodman \& Witten 1985) and 
indirect detection through
observation of energetic neutrinos, gamma rays and other products of
self-annihilation of dark matter particles (Silk \& Srednicki 1984)
.

The event rate for self-annihilation depends quadratically on the local dark
matter density, which falls off with distance from the center of
the halo.
The averaged halo density profile obtained in various numerical simulations 
diverges at the centre but is otherwise smooth and 
is often fit with a(n asymptotically) double power-law 
(Navarro, Frenk \& White 1996, Moore et al
1998). However, a
consensus on the precise
values of the power exponents, the size of the central core and 
the resolution of fine high-density structures are yet to be achieved. 
The fine structures, the {\it caustics},
are inevitable outcomes of the evolution of a collisionless self-gravitating
system described by the Jeans-Vlassov-Poisson equation (for a one-dimensional
numerical
result see Alard \& Colombi 2005). 
Formally, in three dimensions,  the most common caustics are surfaces of zero 
thicknesses over which the density diverges. \footnote{The general theory of 
singularities (Arnol'd, Shandarin\& Zel'dovich 1982)
also predicts singularities on lines and at points. 
Despite the greater concentration of mass 
in these singularities they probably play a less 
important role in the total annihilation rate 
because they contain a considerably smaller amount of mass.
However, this has not been studied in detail.}
However, a maximum cut-off 
to their density is set by the finite non-negligible velocity dispersion 
of dark matter particles. Their density however remains very high and hence
they can be significant for dark matter search experiments (Sikivie \& Ipser
1992, Sikivie et al 1997, Natarajan 2007).
The effect of velocity dispersion 
in the smearing of the caustics is expected to dominate 
over other effects such as particle discreteness which would also smooth the
caustics but to a far lesser degree. Mergers of haloes can also smear out the
caustics substantially and due to this fact we restrict our study to haloes
that have grown by slow and smooth accretion. Nevertheless, caustics are
robust, in that while they may break up into micro-caustics, they remain
in the fine-scale halo substructure and thereby contribute to the general
clumpiness boost of any annihilation signal.

Analytic studies of the formation of haloes and caustics have been carried out
mainly under various simplifying assumptions, such as spherical 
symmetry, self-similarity, and cold and smooth accretion
(Gott 1975, Gunn 1977, Fillmore \& Goldreich 1984, Bertschinger 1985). 
In an Einstein-de Sitter Universe a spherical overdensity
expands and then {\it turns around} to collapse. After collapse and at late
times, the fluid motion becomes self-similar: its form remains
unchanged when its length is re-scaled in terms of the  radius, 
$r_{\rm ta}$, of the shell that is 
currently at  turn-around and is falling onto the
galaxy for the first time. Physically, self-similarity arises because
gravity is scale-free and because mass shells outside the initial
overdensity are also bound and turn around at successively later
times. Self-similar solutions give power-law density profiles
on the scale of the halo which provides an explanation
for the flattening of the rotation curves of  galaxies.
However, on smaller scales the density profile contains many spikes
({\it i.e.}\ caustics) of infinite density. The position and time
of formation of these caustics are among the many properties that have
been established in the framework of the self-similar infall model 
(Fillmore \& Goldreich 1984, Bertschinger 1985).

In reality, dark matter has a small 
velocity dispersion and haloes do suffer 
from major mergers and non-sphericity. 
However, until numerical simulations achieve sufficient
resolution, the self-similar accretion model
provides a useful guideline to haloes
which have not undergone major mergers.

Here, we use the self-similar model of halo formation and a further elaboration 
which includes the velocity dispersion of dark matter 
(Mohayaee \& Shandarin 2006) as a first {\it guidline} to describe the evolution 
of the smallest haloes which have survived major
merger and disruption until
now and have grown only by slow accretion. 

The application of self-similar model to such haloes can be viewed from two
contradictory angles. One might assume that
 minihaloes are expected to be
well-represented by this model, since they contain no substructures, have not
undergone merger and grow very slowly only by smooth mass accretion.
On the other hand, minihaloes are not typical haloes and selfsimilar accretion
model is formulated to describe the evolution of a characteristic halo. 

Keeping both of these issues in mind, we use selfsimilar model only as a first
guidline for the evolution of minihaloes.
A large number of them have been found in
simulations (Diemand et al 2005). The simulations 
estimate the size of these haloes to be of about
0.01 pc (half mass radius) and their mass of $10^{-6}$M$_\odot$ at $z=26$. 
Due to resolution problems, these simulations are stopped at this redshift and
typical evolution of galactic scale haloes is extended to minihaloes and the
conclusion is drawn that  about
$10^{15}$ of these haloes could exist in the halo of MilkWay today.
We assume that at least a
fraction of these haloes have evolved by slow accretion model from
$z=26$ until
now and use the selfsimilar model to evaluate their radius and mass at $z=0$,
which are respectively $1$ pc and $10^{-5}$ M$_\odot$.
For these haloes and working self-consistently within our model  
including the contribution from the caustics, we demonstrate that in the outer
regions of these haloes caustics can boost the annihilation signal by about 40\%.

\section{Secondary-infall model with velocity dispersion }

\begin{figure}
\resizebox{70mm}{!}{\includegraphics{{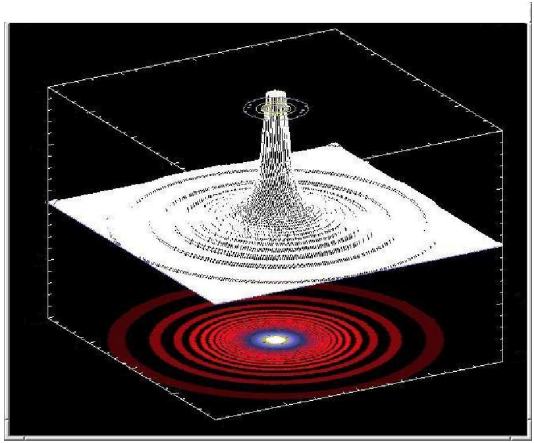}}}
\caption
{
A surface-contour plot of the caustic density.
In the self-similar model, caustics form concentric shells
of increasing density and decreasing thicknesses and separations as we
approach the center of the halo.
}
\label{schematic}
\end{figure}

The haloes considered in this work grow by smooth and slow accretion.
A good example are the earth-mass 
haloes which were recently resolved (at z=26) in
numerical simulations (Diemand et al 2005) and which although small are
expected to have clean spherical caustics.
We expect that at least a fraction of these haloes
have survive disruption and major merger and grow by self-similar accretion model
to a virial radius of about $1$ pc and a mass of about $10^{-5}$ M$_\odot$.
 
To comply with the requirement of slow accretion, we 
fix the value of the parameter $\epsilon$ in
the initial density perturbation
$\delta\sim M_i^{-\epsilon}$,
 where $M_i$ is the initial mass, to unity.
We emphasis that the self-similar model aims at 
describing the evolution of a typical halo. Typical haloes 
have mass variance ($\sigma(M)$ which 
varies as $M^{-(n+3)/6}$, which sets $\epsilon=(n+3)/6$, where $n$ is the power
  spectrum index. A typical $\sigma(M)$
 fluctuation grows as $t^{2/(3\epsilon)}$. Minihaloes correspond to 
the limit $n\rightarrow -3$ part of the spectrum. For this part of the
spectrum, there are mass fluctuations of comparable amplitude on all scales and
consequently adiabatic invariance does not apply for such fluctuations.

Hence, we use the self-similar model only as a first {\it guidline} for the growth of
minihaloes. We assume that minihaloes of mass $10^{-6}M_\odot$ have grown by
very small accretion from $z=26$ to $z=0$. 
Once again, slow accretion corresponds to 
the case of $\epsilon=1$ in the work of Fillmore and 
Goldreich (1984), hence we shall adopt this value for $\epsilon$.

The self-similar density profile is given by (Bertschinger 1985)
\be
{\rho\over\bar\rho}={\pi^2 \over 8 \lambda^2}\sum_j
(-1)^j {\rm exp}\left(-{2\over 3}\xi_j\right)
\left({d\lambda\over d\xi}\right)_j^{-1}
\label{eq:totaldensity}
\ee
where $\bar\rho$ is the critical 
density and
\be
\lambda={r\over r_{ta}}
\ee
is the dimensionless radius and $r$ is the physical radius
and $\xi_i={\rm ln}(t/t_{\rm ta})$ is the dimensionless time
given in terms of the turnaround time, $t_{\rm ta}$, of the particle that is
at the $j$th point where $\lambda=\lambda(\xi)$ (see Bertschinger 1985 for
further explanation). 

\begin{figure}
\resizebox{72mm}{!}{\includegraphics{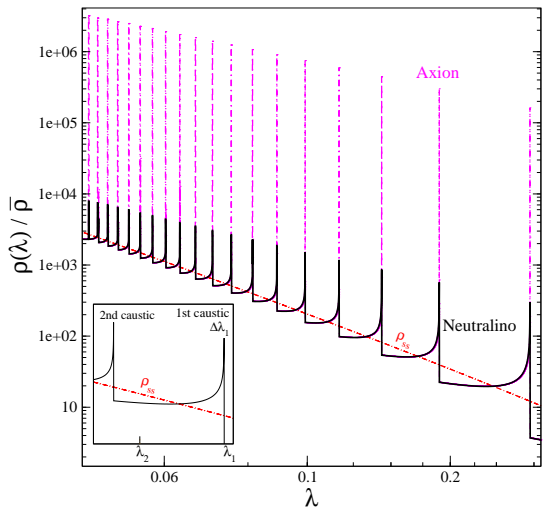}}
\caption
{
The plot is made by numerically solving (\ref{eq:totaldensity})
and cutting the caustics using (\ref{eq:density-caustic}).
The density of halos 
(divided by the critical density $\bar\rho$) can be enhanced significantly at the
caustics. This enhancement is far larger for axions (dotted violet spikes)
than for neutralinos (continuous black line).
The number of the streams increases to the center of the halo which explains
the rapid growth of the smooth component of the density profile and the 
caustic contribution to the halo density diminishes. 
The dashed-dotted red line marked $\rho_{\sf ss}$ is the approximate self-similar 
density profile given by (\ref{density}).
The inset shows a magnified view of the second and third caustics
for neutralinos.
}
\label{fig:caustic-cutoff}
\end{figure}

The density (\ref{eq:totaldensity}) is evaluated numerically and plotted in
Fig.\ \ref{fig:caustic-cutoff} after an appropriate cut-off of the caustics which
shall be discussed now.
In principle the density at the caustics
diverges if the velocity dispersion of dark matter is zero.
In the presence of a small velocity dispersion the maximum density and
thickness of the caustic shells and their density profiles have been
evaluated (Mohayaee \& Shandarin 2006). 
The maximum density at the caustics and their profile are given by  
\be
\rho_{\rm caustic, k} = 
\left\{
\begin{array}{ll}
{ G_k\over \sqrt{\vert \Delta\lambda_k \vert} }\bar\rho 
&\quad \, {\lambda_k-|\Delta\lambda_k| <\lambda< \lambda_k} \;\\
\\
{G_k\over\sqrt{\lambda_k-\lambda}}\bar\rho  &\quad 
{\lambda<\lambda_k-|\Delta\lambda_k|}\\
\end{array}
\right.
\label{eq:density-caustic}
\ee
where $\lambda_k$ is the non-dimensional
radius
of the $k$th caustic counted inwards and
\be
G_k={\pi^2\over 4\sqrt{-2\lambda_k^{\prime\prime}}}{e^{-2\xi_k/3}\over
  \lambda_k^2}
\ee
and the thickness of the caustic shell is given by
\be
\Delta \lambda_k= {(3\pi)^{2/3}e^{5\xi_k/9}\Lambda_k \over 4}\, {t\, \sigma(t)\over 
r_{\rm ta}},
\label{thickness}
\ee
where $t$ is the age of the Universe, $\sigma$ is 
the present-day velocity dispersion 
of dark matter particles which is that at 
decoupling re-scaled with the expansion factor. The values of these
parameters vary from one caustic to another  (see Table 1
of Mohayaee \& Shandarin 2006 for the first ten caustics). 
The profile (\ref{eq:totaldensity}) together with appropriate cut-off given by
(\ref{eq:density-caustic}) is plotted in Fig.\ \ref{fig:caustic-cutoff}.

The peaked density profile given by (\ref{eq:totaldensity}) and
shown in Fig.\ \ref{fig:caustic-cutoff} has to be evaluated
numerically. However, as is evident from  Fig.\ \ref{fig:caustic-cutoff}
a ``self-similar'' profile\footnote{This profile can also be well-fitted by a
  power-law and an exponential cut-off} is reached which we fit with
\be
\rho_{\sf ss}=
{2.8 \lambda^{-9/4}\over (1+\lambda^{3/4})^2}
\bar\rho\,,
\label{density}
\ee
as shown in Fig.\ \ref{fig:caustic-cutoff} by the dashed-dotted red line, marked
$\rho_{\sf ss}$. 
The
turnaround radius, $r_{ta}$ can now be evaluated by considering that at
the virial radius the density is about $200$ 
times the background density,  and is given by
$
r_{\rm ta}\sim 4 r_{\rm vir},
$
which corresponds to the density 
profile given by (\ref{density}).
This approximate profile has been shown to be a good fit also
to the mass profile (see Mohayaee \& Shandarin 2006). In the next 
section we shall show that using this profile which ignores the caustics would
yield an under-estimated value for the flux.

Both the extrapolated numerical and the approximate 
density profiles shown in 
Fig.\ \ref{fig:caustic-cutoff} formally diverge
at the centre. 
However, due to finite dark matter velocity
dispersion, haloes can develop central cores.
Dark matter haloes are expected to have central cores
due to the dark matter velocity dispersion, 
self-annihilations at the centre, angular momentum, tidal 
and various other effects.
The core could be very small and the minimum scale associated with 
a generic dark matter merging history would 
conserve traces of the original cores in the initial substructure.
These should be of order
the free-streaming mass as for example computed in Bertschinger (2006).

In principle for small core sizes the total flux from the whole of the halo 
is dominated by the annihilation in
the centre of the
halo and the boost due to caustics is negligible. However, we shall show in
the next section that
the differential (similarly cumulative) flux would be distinctly marked by 
the caustics and shall have a sawteeth pattern
and the contribution to the total flux from the outer region of these haloes
by the caustics is significant and can yield a boost factor of about $40\%$.

\section{The flux due to self-annihilation including the effect of caustics }

Caustics if detected would be clear evidence of the existence of dark matter
and could rule out alternative models of gravity. Two major methods for their
detection are through gravitational lensing (see e.g. Gavazzi et al 2006) 
and the
flux of dark matter annihilation product which is expected 
to be significantly enhanced by the caustics.
Here we shall discuss the second method.
 
The flux of the self-annihilation 
product ({\it e.g.} $\gamma$-rays) is given by
\be
{\cal F}lux\sim  \int \rho^2 (4\pi r^2)\, dr\,,
\label{F}
\ee
where the proportionality coefficient is a function of
dark matter particle mass, interaction cross section and the number of
photons produced per annihilation.

\begin{figure}
\resizebox{70mm}{!}{\includegraphics{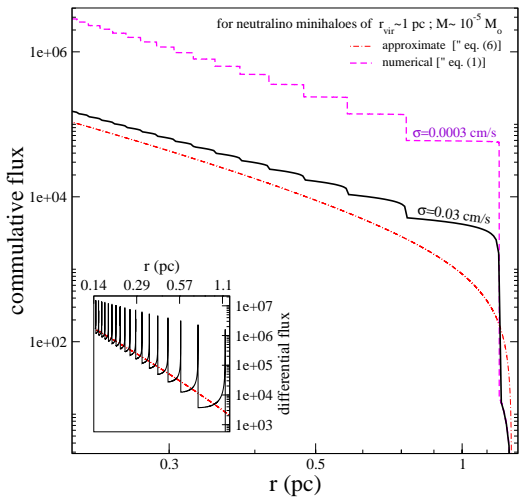}}
\caption
{
Cumulative flux is obtained by summing the flux 
inwards: {\it i.e.}\ from the first
outer caustic towards the most inner ({\it i.e.}\ the integral (\ref{F}) evaluated
 inwards). The flux is shown for 
two different values 
of the velocity dispersion. 
The red dashed line shows the cumulative flux 
obtained by using our approximate
analytic expression for the density (\ref{density}) which neglects the
contribution from the caustics and can considerably underestimate the
annihilation flux
and ignore the distinct sawteeth characteristic of the caustics.
The inset shows the differential flux (integrand of expression (\ref{F}))
using the full density profile (\ref{eq:totaldensity}) as shown by the
solid black spiky line and the approximate profile (\ref{density}) as shown by
the dashed-dotted red line.
The sawteeth pattern is once again neglected in using the later profile. 
}
\label{fig:flux}
\end{figure}

The differential and cumulative flux ({\it i.e.}\ the integrand in 
expression (\ref{F}) and the integral evaluated 
from $r_{\rm ta}$ inwards) for neutralino 
($\sigma=0.03$ cm/s) and a minihalo of 
$r_{\rm ta}=3.24$ pc (which corresponds to a virial radius of about $0.8$ pc) 
is shown in Fig.\ \ref{fig:flux}. 
The fluctuations, due to caustics, become less prominent as we go towards the
centre.  Decreasing the velocity
dispersion would increase both the amplitude of the peaks in the density
profile and the fluctuations in the flux, as 
shown in Figs.~\ref{fig:caustic-cutoff} and  \ref{fig:flux}. 

Using our numerical solution 
to (\ref{eq:totaldensity}) and approximation (\ref{density}),
 we can now determine the flux from the neutralino 
minihaloes (Diemand et al 2005) 
 and its enhancement due to the first twenty caustics. Clearly the total flux 
from the whole halo is dominated by the emission from the centre, where the
density of the caustics reaches the background density (see also . 
However in the
outer regions where the first twenty caustics 
dominate, as shown in Fig.\ 
\ref{fig:caustic-cutoff} the ratio of the flux using 
the self-similar density profile 
given by (\ref{density}) and the complete density profile
(\ref{eq:totaldensity}) gives a boost factor of about
\be
Boost=1.4\,
\ee
Thus, not only we expect a distinct signature on the cumulative and
(similarly differential) flux due to caustics as
highlighted schematically in Fig.\ \ref{schematic} and shown 
numerically in Fig.\ \ref{fig:flux}, we also expect that 
the total flux from the outer halo region including the first twenty caustics to be
boosted by about \%40. Quantitative works on the gamma-ray flux is not carried
out here, as it
requires more realistic model than the self-similar model which can at best explain
the growth of a typical halo. Minihaloes are atypical in the sense that they
evolve in isolation, accreting  almost no mass.

In conclusion, we have modeled dark matter haloes by an extended version of 
secondary infall model        
to include non-vanishing velocity dispersion. 
We have shown that the differential and cumulative fluxes would have distinct
sawteeth pattern due to caustics.
We have demonstrated that
caustics can boost the total annihilation flux by about $40\%$ percents in
the outer regions of smallest haloes of about $10^{-5}$ M$_\odot$. 
As for the prospect of detecting caustics,
the nearest minihaloes could be detectable in gamma rays 
by proper motions observed with
GLAST (Koushiappas 2006), and should display a caustic-like substructure.
One would expect to find a series of caustics, detectable as arclets. The
predicted spacings could be used as a template to dig more deeply into the
noisy background.

{\sf Acknowledgment:} S.S. acknowledges the support from LANL T8, CITA and IAP
during sabbatical year 2005-06 when most of this study was done. We thank M.
Kuhlen and J. Diemand for discussions.




\end{document}